\documentclass[twocolumn,prl,aps,superscriptaddress,showpacs]{revtex4}

\usepackage{bm}\usepackage{graphicx}

\begin{document}

\title{Magnetic order in Ce$_{0.95}$Nd$_{0.05}$CoIn$_{5}$: the $\bf{Q}$-phase at zero magnetic field}

\author{S. Raymond, S. M. Ramos, D. Aoki, G. Knebel, V. Mineev and G. Lapertot}
\affiliation{SPSMS, UMR-E 9001, CEA-INAC/UJF-Grenoble 1, 38054 Grenoble, France}
\date{\today}

\begin{abstract}
We report neutron scattering  experiment results revealing the nature of the magnetic order occurring in the heavy fermion superconductor Ce$_{0.95}$Nd$_{0.05}$CoIn$_{5}$, a case for which an antiferromagnetic state is stabilized at a temperature below the superconducting transition one. We evidence an incommensurate order and its propagation vector is found to be identical to that of the magnetic field induced antiferromagnetic order occurring in the stoichiometric superconductor CeCoIn$_{5}$, the so-called Q-phase. The commonality between these two cases suggests that superconductivity is a requirement for the formation of this kind of magnetic order and the proposed mechanism is the enhancement of nesting condition by $d$-wave order parameter with nodes in the nesting area.

\end{abstract}

\maketitle


The interplay between magnetism and superconductivity is an essential topic whose investigation is common to several families of strongly correlated electron systems: cuprates, iron based superconductors and heavy fermion systems. These systems share the common point that the antiferromagnetic (AFM) ground state can be tuned to a quantum critical point where the N\'eel temperature, $T_N$, reaches zero as a function of pressure, chemical substitution or magnetic field. At this quantum critical point superconductivity often emerges but there is no paradigm: magnetism and superconductivity can coexist or expel each other depending of each system. In this framework, many investigated systems have a N\'eel temperature higher than the superconducting transition temperature, $T_{c}$. The opposite case $T_N < T_c$, known for rare earth transition metal borocarbide compounds and Chevrel phases \cite{Maple}, is much less studied in strongly correlated electron systems especially for itinerant electron ones where the same electrons participate to magnetism and superconductivity. This situation arises certainly from the lack of convenient experimental realization of this scenario. Paradoxically a case where it is nonetheless studied is a complex one: the one of  magnetic field induced AFM order starting from a superconducting ground state. This case is common to the cuprate La$_{2-x}$Sr$_{x}$CuO$_{4}$, showing both field induced or field enhanced magnetism \cite{Chang}, and to the heavy fermion compounds CeRhIn$_{5}$ under pressure and CeCoIn$_{5}$ at ambient pressure \cite{Knebel}. This kind of cooperative effect between magnetism and superconductivity reaches its pinnacle in CeCoIn$_{5}$ where the field induced AFM phase disappears when superconductivity is suppressed at the upper critical field $H_{c2}$.

CeCoIn$_{5}$ has the highest superconducting transition temperature among Ce heavy fermion compounds ($T_c$ = 2.3 K) \cite{Petrovic}. It crystallizes in a tetragonal structure (space group
P4/mmm) and the superconducting gap symmetry is considered to be the singlet $d_{x2-y2}$ state \cite{revue1}. A field induced ordered phase (FIOP) occurs for a magnetic field applied in the basal plane of the tetragonal structure, in a narrow range of temperature and magnetic field below 300 mK and above 10.5 T, the upper critical field being 11.4 T for this geometry. It was shown by Kenzelmann et al. that the FIOP is an AFM phase: the corresponding order is incommensurate with a propagation vector $\bf{k_{IC}}$=(0.45, 0.45, 0.5) and a magnetic moment of 0.15 $\mu_{B}$ aligned along the $c$-axis of the tetragonal structure; the FIOP was thereafter named Q-phase \cite{Kenzelmann}. Prior to this microscopic finding, the FIOP was thought to be the realization of a modulated superconducting phase, the so-called Fulde-Ferrell-Larkin-Ovchinnikov (FFLO) state \cite{revue2}. Despite the finding of magnetic order, this possibility cannot be excluded and a multicomponent order parameter may exist:  the FIOP of CeCoIn$_{5}$ might therefore be a coupling of superconducting and magnetic order parameters. The uniqueness of the interpenetrated magnetic and superconducting properties of CeCoIn$_{5}$ turns out to be a playground for advanced condensed matter physics concepts leading to many experimental and theoretical works.

In the present study, we choose another route to reach a case $T_N < T_c$ in a related situation achieved by Nd substitution on CeCoIn$_{5}$. For 0.05 $\leq x \leq$ 0.15, it was shown by bulk measurements that antiferromagnetism occurs in Ce$_{1-x}$Nd$_{x}$CoIn$_{5}$ with still high superconducting transition temperature in the range 1-2 K \cite{Hu}. For higher Nd substitution, superconductivity disappears for $x >$ 0.2, and heavy fermion behavior is superseded by local moment magnetism for $x >$ 0.5. In the present paper, we report that the magnetic order occurring at low temperature inside the superconducting phase for the selected compound Ce$_{0.95}$Nd$_{0.05}$CoIn$_{5}$ ($T_c$ $\approx$ 1.8 K and $T_N$ $\approx$ 0.9 K) has the same propagation vector as the one induced by magnetic field in pure CeCoIn$_{5}$.  We suggest that this magnetic order occurring with $T_N < T_c$ is stimulated by $d$-wave nodal superconductivity.

The neutron scattering measurements were carried out on the cold neutron three axis spectrometer (TAS) IN12 located at ILL, Grenoble. 
The initial neutron beam is provided by a double focusing pyrolitic graphite (PG) monochromator. Higher order contamination is removed before the monochromator by a velocity selector. Diffraction measurements were carried out with the horizontally focusing PG analyzer used for reducing the background. The spectrometer was setup in long-chair configuration with open-open-open collimations. The search for magnetic Bragg peaks and the order parameter temperature dependence measurement were performed using neutrons of wave-vector $k_i=k_f$ = 1.3 $\AA^{-1}$ while the Bragg peaks rocking curves were collected using  $k_i=k_f$= 1.8 $\AA^{-1}$. The first configuration allows to reduce the incoherent background and the second one allows to reach higher scattering angle and also to reduce neutron absorption. The sample grown by the self-flux method \cite{Canfield} is a single crystal of approximate size 4*7*0.3 mm$^3$and a total mass of 62 mg. It was mounted in a helium-3 cryostat with the [1, -1, 0] axis vertical, the scattering plane being thus defined by [1,1,0] and [0,0,1].
The whole sample used for neutron scattering experiment was characterized by EDX microanalysis which reveals a single homogeneous composition. The specific heat, $C$, of a small part cut from this sample was measured in a standard PPMS apparatus. $C/T$ is shown in Figure 1 as a function of temperature. 
\begin{figure}
\centering
\includegraphics[width=8cm]{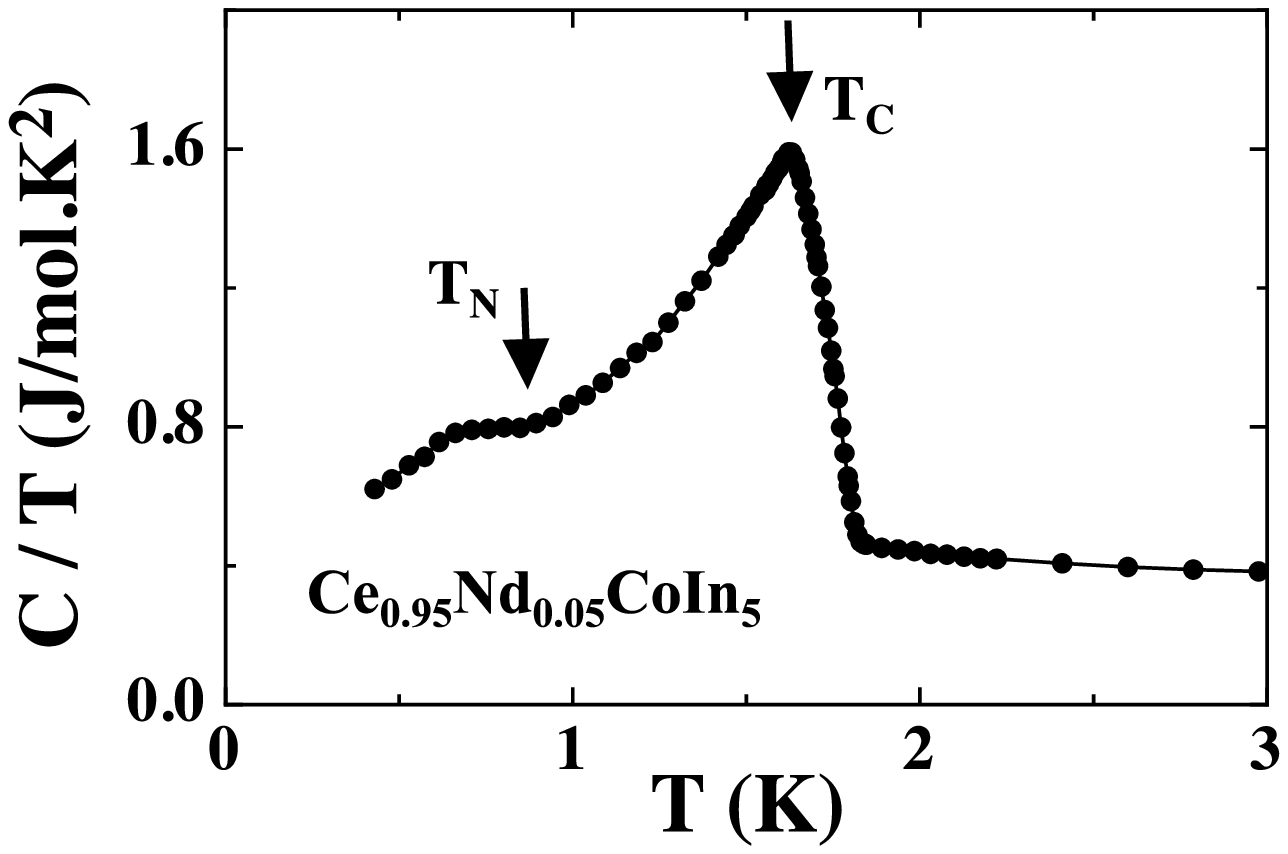}
\caption{Specific heat of Ce$_{0.95}$Nd$_{0.05}$CoIn$_{5}$ divided by temperature as a function of temperature. The arrows indicate the superconducting ($T_c$) and antiferromagnetic ($T_N$) transitions.}
\end{figure}
One can observe two transitions: the first one at 1.84 K and, by further reducing the temperature, the onset of the second one at  0.9 K. The known phase diagram of Ce$_{1-x}$Nd$_{x}$CoIn$_{5}$ \cite{Hu} allows (i) to ascribe the first transition to superconductivity and the second one to antiferromagnetism and (ii) to check the good agreement between $x$, $T_N$ and $T_c$ for samples prepared by two different groups.

The aim of the present work is to investigate the characteristics of the AFM phase occurring inside the superconducting one. In this paper, the scattering vector $\bf{Q}$ is decomposed into $\bf{Q}$=${\bm \tau}$+$\bf{k}$, where ${\bm \tau}$ is a Brillouin zone center and $\bf{k}$ is the propagation vector for a given magnetic structure. The cartesian coordinates, $H$ and $L$, of the scattering vector $\bf{Q}$ are expressed in reciprocal lattice unit (r.l.u.) (($\bf{Q}$=($H$, $H$, $L$)). The search for magnetic signal was performed at low temperature along the lines ($H$, $H$, 0.5) for 0.4 $\leq H \leq$ 1 and (0.5, 0.5, $L$) for 0 $\leq L \leq$ 1. All the reported propagation vectors in compounds related to CeCoIn$_{5}$ are located on these lines. A detail of the scan performed along  ($H$, $H$, 0.5) is shown in Fig. 2a: for $T$=0.37 K (full circles), two peaks are observed at $H$=0.448 $\pm$ 1 r.l.u. and respectively $H$=0.548 $\pm$ 1 r.l.u. These peaks disappear at 1.1 K (empty circles). A scan on the peak at $H$=0.55 along the $L$ direction is shown in Fig. 2b at low temperature. The intensity is maximum for $L$=0.5. No other magnetic peaks have been found; we conclude that the propagation vector of the antiferromagnetic phase investigated is $\bf{k_{IC}}$=(0.45, 0.45, 0.5). 
\begin{figure}
\centering
\includegraphics[width=8cm]{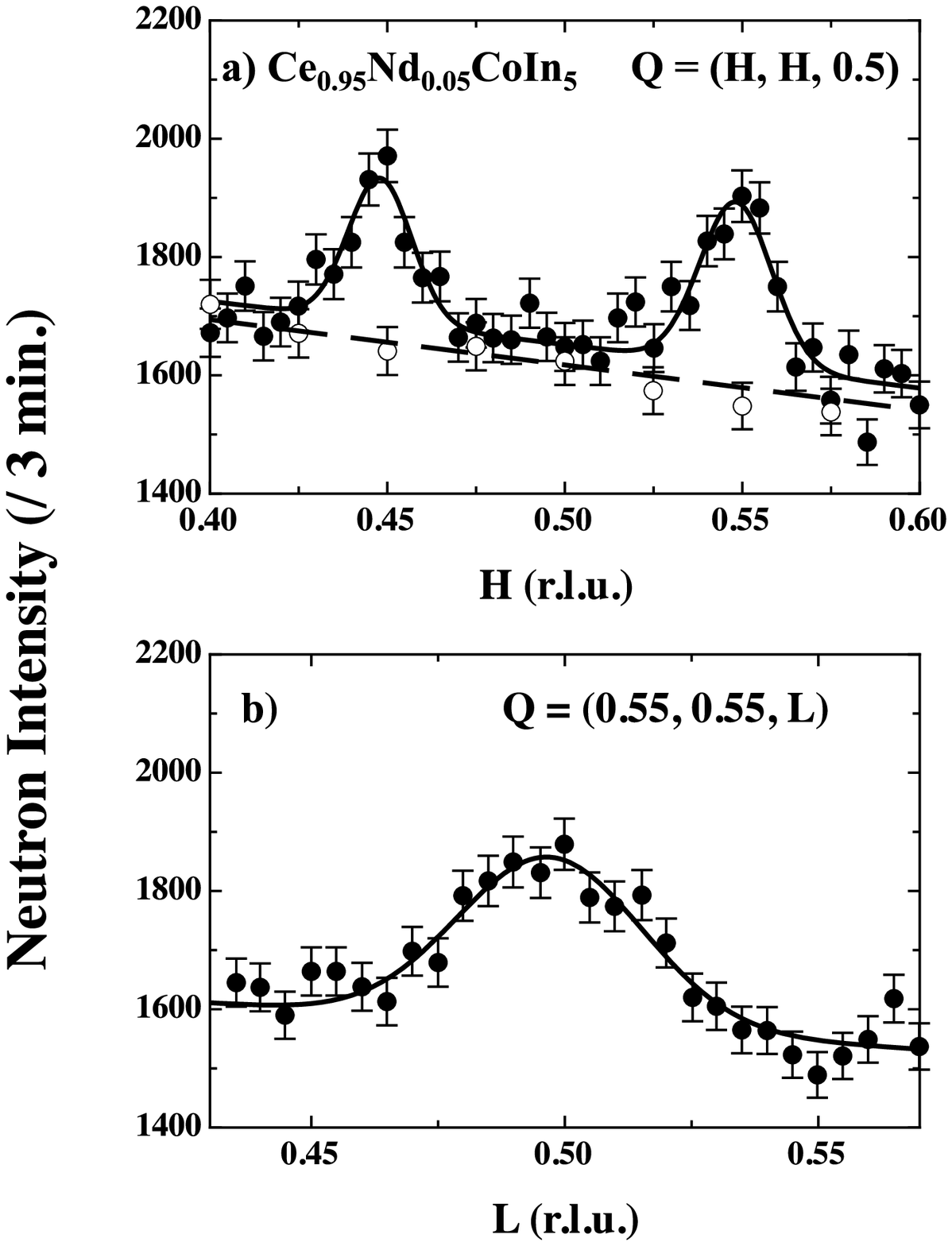}
\caption{$\bf{Q}$ scans performed a) along  ($H$, $H$, 0.5) and b) (0.55, 0.55, $L$). Full (empty) circles are data obtained below (above) $T_N$ at 0.37 K (1.1 K). Lines are gaussian fit with linear background.}
\end{figure}
The temperature variation of the maximum intensity measured at $\bf{Q}$=(0.55, 0.55, 0.5)=(1,1,1)-$\bf{k_{IC}}$ is shown in Figure 3. Only a small number of points were collected due to the weakness of the magnetic signal.  The N\'eel temperature is located between 0.8 and 0.9 K, in agreement with the data from specific heat measurements. To get insight into the magnetic structure, rocking curves measurements were carried out for five magnetic reflections (4 independent ones). Magnetic form factor, Lorentz factor and absorption corrections were taken into account. The two magnetic reflections collected at high scattering angle, (1.45, 1.45, 0.5) and (0.45, 0.45, 2.5), with dominant component of the scattering vector in the plane or respectively along the $c$-axis indicate that the ordered moment is neither solely along the $c$-axis nor confined to the plane (Such characteristic Bragg reflections are often used to obtain the moment direction owing to the fact that neutron scattering probes only magnetism perpendicular to $\bf{Q}$). The use of a cold TAS for performing diffraction experiments allows to detect a weak magnetic signal but does not allow for a full structure determination due to the low number of available Bragg peaks in the scattering plane. Also only one magnetic $\bf{k}$-domain can be investigated in the scattering plane normal to [1, -1, 0]: there is no access to Bragg reflections corresponding to the second domain with the propagation vector  $\bf{k'_{IC}}$=(0.45, -0.45, 0.5). The two magnetic Bragg peaks at $\bf{Q}$=(0.55, 0.55, $\pm$ 0.5) probe almost equally in plane and $c$-axis contributions of the magnetic moment. By normalizing their intensity to the weak intensity nuclear reflection (1,1,0) and the modest intensity nuclear reflection (1,1,1)\cite{Remark}, it is possible, without any hypothesis on the moment direction, to deduce that the ordered moment is bound in the range 0.04 - 0.08 $\mu_{B}$. Hence the higher boundary for the ordered moment in Ce$_{0.95}$Nd$_{0.05}$CoIn$_{5}$ corresponds to half the magnetic moment found in the FIOP at 60 mK and 11 T \cite{Kenzelmann}.

\begin{figure}
\centering
\includegraphics[width=8cm]{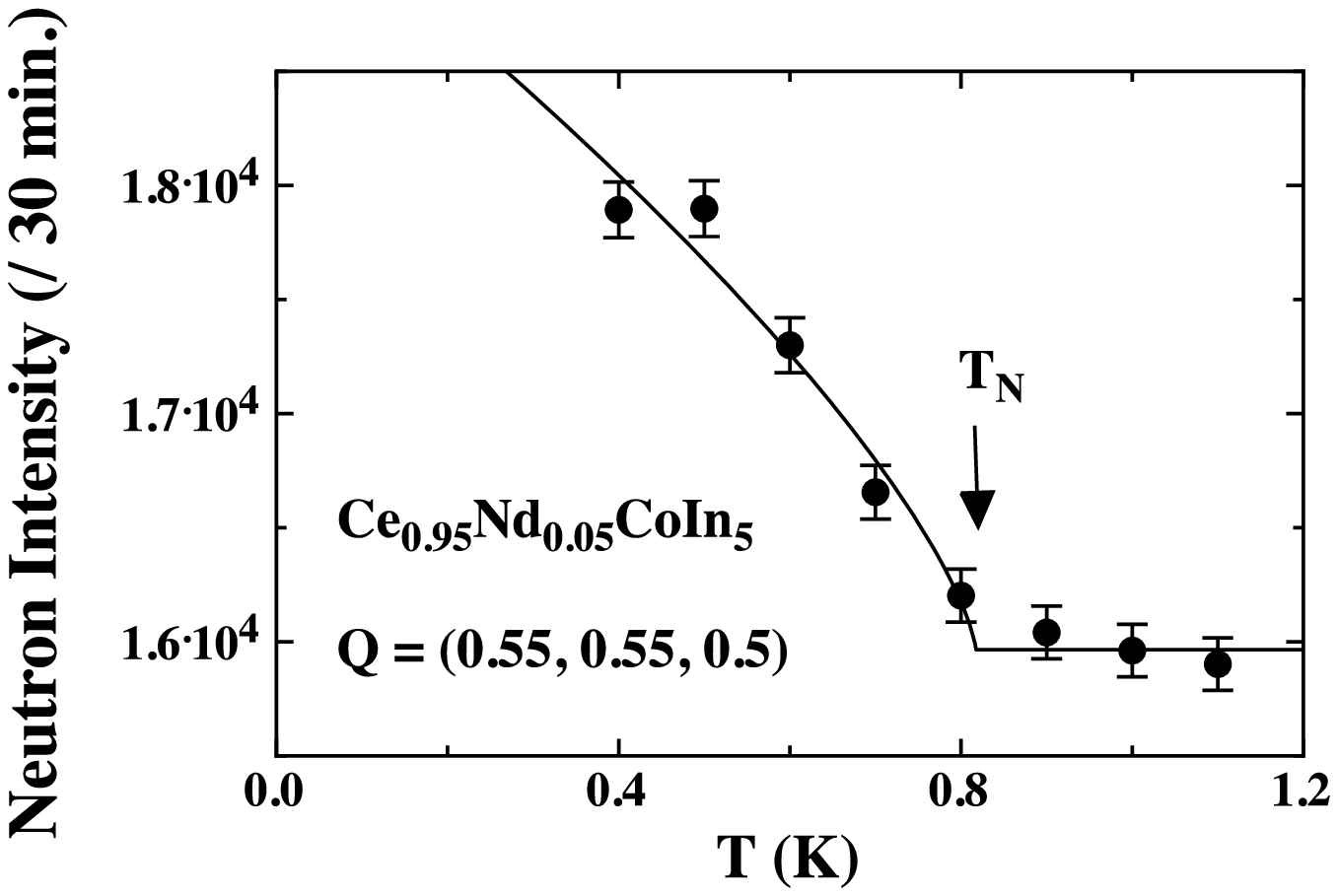}
\caption{Temperature dependence of the neutron intensity measured at $\bf{Q}$=(0.55, 0.55, 0.5). The line is a guide for the eyes.}
\end{figure}

\bigskip
Among the different chemical substitutions leading to magnetic order in CeCoIn$_{5}$, two cases were investigated to our knowledge by neutron diffraction. Antiferromagnetism is observed in CeCo(In$_{1-x}$Cd$_{x}$)$_{5}$ for $x \ge$ 0.075 and superconductivity is found up to $x$=0.125. Neutron diffraction was performed for $x$=0.1 ($T_N$ $\approx$ 3 K, $T_c$ $\approx$ 1.3 K) \cite{Nicklas} and $x$=0.075 ($T_N$ $\approx$ 2.4 K, $T_c$ $\approx$ 1.7 K) \cite{Nair}. Antiferromagnetism with $\bf{k_{AF}}$=(1/2, 1/2, 1/2) is found. In CeCo$_{1-x}$Rh$_{x}$In$_{5}$, for $x \ge 0.2$ antiferromagnetism is induced while superconductivity is still present up to $x$ $\approx$ 0.7. A diffraction experiment performed on a sample with $x$=0.4 with $T_N$ $\approx$ 2.9 K and $T_c$ $\approx$ 1.4 K also reveals a commensurate order with $\bf{k_{AF}}$ \cite{Ohira} while another group reports in this concentration range the coexistence of ordering with $\bf{k_{AF}}$ and another  incommensurate phase with $\bf{k'}$ =(0.5, 0.5, 0.42) \cite{Yokoyama}. A likely reason for this discrepancy is a sample dependence in relation with slight changes in impurities and defects \cite{Discussion}. For completeness, it should be mentioned that CeRhIn$_{5}$ orders with a propagation vector (0.5, 0.5, 0.31) and that the $c$-axis incommensuration shifts to $\approx$0.4 in the vicinity of the pressure where superconductivity is induced \cite{Raymond,Aso}. This case would correspond to a rich Rh content of CeCo$_{1-x}$Rh$_{x}$In$_{5}$ and certainly to a different physics from that  of the low Rh content. This distinction between the physics at low and high Rh substitution is supported by the Fermi-surface reconstruction occurring for $x$ $\approx$ 0.6 \cite{Goh}. From all these data, it appears that commensurate antiferromagnetism is favored when slight doping drives CeCoIn$_{5}$ toward a magnetic phase and superconductivity still exists with $T_N > T_c$. 
In contrast, the magnetic ordering  found in the present study for Ce$_{0.95}$Nd$_{0.05}$CoIn$_{5}$ is incommensurate with incommensuration in the tetragonal basal plane and surprisingly the propagation vector is the one of the FIOP of pure CeCoIn$_{5}$. This finding has strong consequences on theoretical works describing the  FIOP of CeCoIn$_{5}$ since it shows that the propagation vector is not determined by the magnetic field itself but is related to an incipient magnetic instability revealed here by Nd substitution. This magnetic order is unique among 1-1-5 Ce compounds. The key point is certainly that Ce$_{0.95}$Nd$_{0.05}$CoIn$_{5}$ and the FIOP share the common feature that the AFM phase occurs inside the superconducting phase ($T_N < T_c$).  

From this similarity and the fact that all other studied CeCoIn$_{5}$ alloys have commensurate AFM structure for low substitution, we suggest that superconductivity is the key ingredient to realize the incommensurate magnetic ground state with $\bf{k_{IC}}$. In order to pinpoint the possible mechanism involved, it is useful to review the theories for the magnetic field induced AFM phase of CeCoIn$_{5}$. The most common propositions to explain the FIOP include some of the following elements: (i) multicomponent order parameter with a coupling between
spin density wave and among other possibilities FFLO
superconductivity \cite{Yanase}, $\pi$-wave superconductivity \cite{Aperis},
pair density wave \cite{Agterberg} (ii) the role played by the vortex
lattice \cite{Suzuki} (iii) the importance of Pauli limited superconductivity \cite{Ikeda} (iv) spin-exciton condensation \cite{Michal} (v) the
enhancement of nesting by superconductivity \cite{Kato}.  Our
new result favors the last mechanism since it is transposable to a situation without magnetic field and hence to our findings for Ce$_{0.95}$Nd$_{0.05}$CoIn$_{5}$. The stimulation of antiferromagnetic instability by $d$-wave nodal superconductivity  was considered in two dimensions by Y.  Kato, C. D. Batista and I. Vekhter \cite{Kato}. In this model, the nesting conditions are created by the magnetic field induced elliptical Fermi pockets of spin-polarized Bogoliubov quasiparticles. This mechanism does not work in the absence of magnetic field. However, the corresponding underlying concept transposable to zero magnetic field is that the magnetic ordering originates from the enhancement of the nesting condition by $d$-wave superconductivity with node in the nesting area: the imperfect normal state nesting between convex and concave parts of the Fermi surface connected by the three dimensional wave-vector $\bf{k_{IC}}$ is improved in the superconducting state having nodes connected by $\bf{k_{IC}}$. In other words, the presence of nodes in the quasiparticle spectrum at the quasi-nesting wave-vector makes the regions of gapped Fermi surface near the quasi-nesting area more flat than in the normal state. So far, the existence of almost nested parts of the Fermi surface connected by  ${\bf k}_{IC}$ is not specifically pointed out in band structure calculation, dHvA and ARPES measurements  \cite{Settai, Maehira, Koi} and this deserves further investigations.  Since in pure CeCoIn$_{5}$, the superconducting state is not sufficient to stimulate the AFM phase, we can speculate that the magnetic field provides further fine tuning. It cannot be excluded that additional mechanisms are also involved as well.


In summary, we evidence that the magnetic order occurring for slightly Nd doped CeCoIn$_{5}$ has the same incommensuate propagation vector as the magnetic field induced AFM phase of pure CeCoIn$_{5}$. 
Since in both cases $T_N < T_c$ and since all the other related systems with $T_c < T_N$ have commensurate antiferromagnetic order, this suggests that superconductivity is required to reach this peculiar ground state.
The proposed mechanism is the enhancement of nesting condition by superconductivity for order parameter having nodes connected by the quasi-nesting vector.
This finding opens the door for further investigations of the cooperative effects between magnetism and superconductivity in CeCoIn$_{5}$ related systems as well as asks for sharpening the role played by the magnetic field in the FIOP.

\bigskip

We acknowledge K. Mony for sample preparation and F. Bourdarot and J.-P. Brison for helpful discussions. Cerium was provided by the Materials Preparation Center, Ames Laboratory, US DOE Basic Energy Sciences, Ames, IA, USA, available from: www.mpc.ameslab.gov.


\begin{thebibliography}{99}

\bibitem{Maple} M.B. Maple, E.D Bauer, V.S. Zapf and J. Wosnitza in Superconductivity, K.H Bennemann and J.B. Ketterson eds, Springer 2008, p641 and references therein.
\bibitem{Chang} J. Chang, C. Niedermayer, R. Gilardi, N.B. Christensen, H.M. R$\o$nnow, D.F. McMorrow, M. Ay, J. Stahn, O. Sobolev, A. Hiess, S. Pailhes, C. Baines, N. Momono, M. Oda, M. Ido and J. Mesot, Phys. Rev. B $\bf{78}$, 104525 (2008) and references therein.
\bibitem{Knebel} G. Knebel, D. Aoki and J. Flouquet, C.R. Physique $\bf{12}$, 542 (2011) and references therein.
\bibitem{Petrovic} C. Petrovic, P.G. Pagliuso, M.F. Hundley, R. Movshovich, J.L. Sarrao, J.D. Thompson, Z. Fisk and P. Monthoux, J. Phys. Condens. Matter $\bf{13}$, L337 (2001).
\bibitem{revue1} J. L. Sarrao and J. D. Thompson: J. Phys. Soc. Jpn. $\bf{76}$ (2007) 051013 and references therein.
\bibitem{Kenzelmann} M. Kenzelmann, Th. Str\"assle, C. Niedermayer, M. Sigrist, B. Padmanabhan, M. Zolliker, A.D. Bianchi, R. Movshovich, E.D. Bauer, J.L. Sarrao and J.D. Thompson, Science $\bf{321}$, 1652-1654 (2008).
\bibitem{revue2} Y. Matsuda and H. Shimahara: J. Phys. Soc. Jpn. 76 (2007) 051005 and references therein.
\bibitem{Hu} R. Hu, Y. Lee, J. Hudis, V.F. Mitrovic and C. Petrovic, Phys. Rev. B $\bf{77}$, 165129 (2008).
\bibitem{Canfield} P.C. Canfield and Z. Fisk, Phil. Mag. B $\bf{65}$, 1117 (1992). 
\bibitem{Nicklas} M. Nicklas, O. Stockert, T. Park, K. Habicht, K. Kiefer, L.D. Pham, J.D. Thompson, Z. Fisk and F. Steglich, Phys. Rev. B $\bf{76}$, 052401 (2007).
\bibitem{Remark} The intensity of the two nuclear reflections (1,1,0) and (1,1,1) can be described with the same scale factor. Therefore no extinction correction is required probably due to the very small size of our sample.
\bibitem{Nair} S. Nair, O. Stockert, U. Witte, M. Nicklas, R. Schedler, K. Kiefer, J.D. Thompson, A.D. Bianchi, Z. Fisk, S. Wirth and F. Steglich, PNAS 107, 9537 (2010).
\bibitem{Ohira} S. Ohira-Kawamura, H. Shishido, A. Yoshida, R. Okazaki, H. Kawano-Furukawa, T. Shibauchi, H. Harima and Y. Matsuda, Phys. Rev. B $\bf{76}$, 132507 (2007).
\bibitem{Yokoyama} M. Yokoyama, N. Oyama, H. Amitsuka, S. Oinuma, I. Kawasaki, K. Tenya, M. Matsuura, K. Hirota and T.T. Sato, Phys. Rev. B $\bf{77}$, 224501 (2008).
\bibitem{Discussion} S. Ohira-Kawamura, H. Kawano-Furukawa, H. Shishido, R. Okazaki, T. Shibauchi, H. Harima and Y. Matsuda, Phys. Status Solidi A $\bf{206}$ 1076 (2009).
\bibitem{Raymond} S. Raymond, G. Knebel, D. Aoki and J. Flouquet, Phys. Rev. B $\bf{77}$, 172502 (2008).
\bibitem{Aso} N. Aso, K. Ishii, H. Yoshizawa, T. Fujiwara, Y. Uwatoko, G.-F. Chen, N.K. Sato and K. Miyake, J. Phys. Soc. Japan $\bf{78}$, 073703 (2009).
\bibitem{Goh} S.K. Goh, J. Paglione, M. Sutherland, E.C.T. O'Farrell, C. Bergemann, T.A. Sayles and m.B. Maple, Phys. Rev. Lett. $\bf{101}$, 056402 (2008).
\bibitem{Settai} R. Settai, H. Shishido, S. Ikeda, Y. Murakawa, M. Nakashima, D. Aoki, Y. Haga, H. Harima and Y. \"{O}nuki, J. Phys. Condens. Matter $\bf{13}$, L627 (2001). 
\bibitem{Maehira} T.Maehira, T. Hotta, K. Ueda and A. Hasegawa, J.Phys. Soc. Japan {\bf 72}, 854 (2003).
\bibitem{Koi} A. Koitzsch, I. Opahle, S. Elgazzar, S.V. Borisenko, J. Geck, V.B. Zabolotnyy, D. Inosov, H. Shiozawa, M. Richter, M. Knupfer, J. Fink, B. B\"uchner, E.D. Bauer, J.L. Sarrao and R. Follath, Phys. Rev. B $\bf{79}$, 075104 (2009).
\bibitem{Yanase} Y. Yanase and M. Sigrist, J. Phys. Soc. Japan $\bf{80}$, 094702 (2011).
\bibitem{Aperis} A. Aperis, G. Varelogiannis and P.B. Littlewood, Phys. Rev. Lett. $\bf{104}$, 216403 (2010).
\bibitem{Agterberg} D.F. Agterberg, M. Sigrist and H, Tsunetsugu, Phys. Rev. Lett. $\bf{102}$, 207004 (2009). 
\bibitem{Suzuki} K. M. Suzuki, Y. Tsutsumi, N. Nakai, M. Ichioka and K. Machida, J. Phys. Soc. Japan $\bf{80}$, 123706 (2011).
\bibitem{Ikeda} R. Ikeda, Y. Hatakeyama and K. Aoyama, Phys. Rev. B $\bf{82}$, 060510(R) (2010). 
\bibitem{Michal} V.P. Michal and V.P. Mineev, Phys. Rev. B $\bf{84}$, 052508 (2011).
\bibitem{Kato} Y. Kato, C.D. Batista and I. Vekhter, Phys. Rev. Lett. $\bf{107}$, 096401 (2011).






\end{thebibliography}
\end{document}